\documentclass[a4paper, 11pt]{article}

\usepackage[T1]{fontenc}
\usepackage[utf8]{inputenc}
\usepackage{sourceserifpro}
\usepackage{sourcesanspro}
\usepackage{microtype}
\usepackage{graphicx}
\usepackage{booktabs}
\usepackage{array}
\usepackage[a4paper,margin=2.4cm]{geometry}
\usepackage{amssymb}
\usepackage{pifont}
\usepackage{setspace}
\usepackage{natbib}
\usepackage{url}
\usepackage{xcolor}
\usepackage{placeins}
\usepackage{float}
\usepackage{enumitem}
\usepackage{subcaption}
\usepackage{parskip}
\usepackage[nobottomtitles*]{titlesec}

\definecolor{vnavy}{RGB}{14, 38, 84}
\definecolor{vpink}{RGB}{236, 44, 102}

\usepackage[colorlinks=true,linkcolor=black,citecolor=black,urlcolor=black,pdfborder={0 0 0}]{hyperref}
\titleformat{\section}{\sffamily\bfseries\Large}{\thesection}{0.7em}{}
\titlespacing*{\section}{0pt}{22pt}{8pt}
\titleformat{\subsection}{\sffamily\bfseries\large}{\thesubsection}{0.6em}{}
\titlespacing*{\subsection}{0pt}{14pt}{5pt}

\setlist[itemize]{itemsep=3pt,topsep=4pt,leftmargin=1.6em}
\newcommand{\cmark}{\ding{51}}
\newcommand{\xmark}{\ding{55}}
\newcommand{\hd}[1]{\textbf{#1}}
\newcommand{\tb}[1]{\par\noindent{\sffamily\bfseries #1}: }
\newcommand{\framedimg}[2]{{\setlength{\fboxsep}{0pt}\setlength{\fboxrule}{0.5pt}%
  \fcolorbox{gray!60}{white}{\includegraphics[width=#1]{#2}}}}
\newcommand{\shot}[4]{\begin{figure}[H]\centering
  \framedimg{#4\textwidth}{#1}\caption{#2}\label{#3}\end{figure}}

\begin{document}

% ---------- Title banner (first page only) ----------
\renewcommand{\thefootnote}{\fnsymbol{footnote}}
\begingroup
\setlength{\fboxsep}{20pt}
\noindent\colorbox{vnavy}{\parbox{\dimexpr\textwidth-2\fboxsep\relax}{%
  \vspace{4pt}
  {\sffamily\bfseries\fontsize{84}{88}\selectfont\textls[30]{\textcolor{vpink}{VIR}\textcolor{white}{ENA}}\par}
  \vspace{10pt}
  {\color{white}\sffamily\large Virtual Arena for Research, Education, and Democratic Innovation\footnotemark[1]\par}
  \vspace{2pt}}}
\endgroup

\vspace{14pt}
{\sffamily\large\bfseries\color{vnavy} Emma Hoes\footnotemark[2]\quad\textcolor{vpink}{/}\quad K.~Jonathan Klüser\quad\textcolor{vpink}{/}\quad Fabrizio Gilardi\par}
\vspace{2pt}
{\sffamily\color{gray} Department of Political Science, University of Zurich\par}
{\sffamily\small\color{gray}3 October 2026\par}
\footnotetext[1]{This version (v3) is a full rewrite. It describes the current platform, including the Instagram, Facebook, Reddit, X, WhatsApp, and Messenger interfaces, and documents the setup and use of experiments. VIRENA is under active development and currently in use at the University of Zurich, partly supported by the DIZH Innovation Program: 2nd Founder-Call. This preprint will be updated as new features are released. For the latest version and to inquire about demos or pilot collaborations, contact the authors.}
\footnotetext[2]{Corresponding author: \href{mailto:hoes@ipz.uzh.ch}{hoes@ipz.uzh.ch}}
\setcounter{footnote}{0}
\renewcommand{\thefootnote}{\arabic{footnote}}

% ---------- Abstract ----------
\vspace{16pt}
\begingroup
\setlength{\fboxsep}{12pt}
\noindent\colorbox{vnavy!7}{\parbox{\dimexpr\textwidth-2\fboxsep\relax}{%
{\sffamily\bfseries\color{vnavy}\large Abstract}\par\vspace{4pt}
\small Digital platforms shape how people communicate, deliberate, and form opinions. Studying these dynamics has become harder because of restricted data access, ethical limits on real-world experiments, and the technical demands of existing research tools. VIRENA (Virtual Arena) is a platform for controlled experiments in realistic social media environments. Several participants can interact at the same time in replicas of feed-based platforms (Instagram, Facebook, Reddit, X) and messaging apps (WhatsApp, Messenger). AI agents powered by large language models (LLMs) join the participants with configurable personas and human-like timing. Researchers set up experiments in a visual interface without programming: they define conditions, schedule stimulus content, add moderation rules, assign participants randomly to conditions, and export the data. VIRENA supports designs that were hard to run before, such as studying human--AI interaction in realistic social settings, comparing moderation interventions, and observing group deliberation as it unfolds. Participants and data stay within institutional control, and the platform links to survey and recruitment tools. This paper describes how VIRENA works and how to use it.\par}}
\endgroup

\newpage

\section{Introduction}

\subsection{The Challenge of Studying Digital Communication}

Social media platforms are central to public discourse, yet studying how people behave and are influenced on them has become difficult. Major platforms have restricted access to data and APIs, which makes independent study of information spread, polarization, and online behavior hard \citep{bruns2019, freelon2018, wagner2023}. Experiments on real platforms raise further problems: informed consent is difficult to obtain, participants may see harmful content, their digital footprints may change, and researchers cannot control the environment.

\subsection{VIRENA in Brief}

VIRENA is a research platform that combines features usually found in separate tools:

\tb{Platform-specific interfaces}Each simulation copies the look and the interaction options of one platform: Instagram, Facebook, Reddit, X, WhatsApp, or Messenger.

\tb{Live multi-user interaction}Several real participants share one simulation and respond to each other in real time.

\tb{Configurable AI agents}LLM-powered agents post, comment, and chat alongside participants. Researchers define their persona, timing, and triggers.

\tb{Moderation as a research variable}Researchers choose how content is detected (keywords, patterns, AI prompt) and what happens (flag, delete, popup warning).

\tb{Scripted content}Posts, comments, and messages appear at exact times.

\tb{Group-level randomization}Groups that gather in a waiting room are assigned randomly and evenly to conditions.

\tb{No code}Everything is configured in a web interface.

\tb{Survey and recruitment integration}Links, external participant IDs, and redirect URLs connect VIRENA to tools such as Qualtrics, QuestionPro, and Prolific.

\subsection{Status and Access}

VIRENA is in use for research at the University of Zurich. Researchers who want a guided demo, a pilot study with their own design, or a collaboration can contact the authors.

\subsection{Paper Overview}

Section~\ref{sec:related} reviews related work. Section~\ref{sec:overview} gives an overview and defines the key terms. Section~\ref{sec:guide} is a user guide that follows the order in which a study is built and run. Section~\ref{sec:platforms} describes the platform-specific interfaces. Section~\ref{sec:usecases} presents use cases, and Section~\ref{sec:discussion} discusses limitations and next steps.

\section{Related Work}\label{sec:related}

\subsection{Single-Participant Simulations}

The Truman Platform \citep{difranzo2018} gives each participant a realistic social media experience in which all other ``users'' are scripted bots. Every participant sees the same content, which gives strong experimental control. Participants never meet real people, and the bots follow scripts. SocialSim \citep{agha2025} adds direct messaging, but the single-participant design remains.

\subsection{Independent Feed-Based Research Platforms}

\citet{neubrander2026}, building on a platform also used by \citet{vanloon2025} and \citet{allamong2025}, built an app in which participants see a feed filled with LLM-powered synthetic users. It allows researchers to manipulate reaction buttons and social norm messages and has supported randomized trials with over 2,000 participants. Each participant still interacts alone with agents. The interface is one generic feed, and moderation is not configurable.

\subsection{Real-Time Multi-User Platforms}

Slurk \citep{schlangen2018} and hivelab \citep{scurrell2026} support groups in text chat with bots, and hivelab randomizes parallel group conversations. Empirica \citep{almaatouq2021} is a framework for real-time multi-user experiments that requires programming. Deliberate Lab \citep{qian2026} is an open-source, no-code platform for synchronous experiments with humans and LLM agents, including agents as moderators. These tools do not replicate specific social media platforms, feeds with images and videos, or platform-specific reactions.

\subsection{Agent-Based Simulations}

OASIS \citep{yang2024} simulates social media dynamics with up to one million LLM agents. This allows the study of information spread, group polarization, and herd effects. It is a computational model without human participants.

\subsection{Educational Games}

Inoculation games such as Bad News \citep{roozenbeek2019}, Harmony Square, and Cranky Uncle teach players to recognize misinformation techniques. They are effective media literacy tools, but they are single-player games with fixed scenarios, not platforms for controlled experiments.

\subsection{The Gap VIRENA Fills}

Most existing tools also require programming skills to configure and deploy. VIRENA combines capabilities that are spread across different tools (Table~\ref{tab:comparison}).

\begin{table}[!htbp]
\centering
\small
\renewcommand{\arraystretch}{1.25}
\resizebox{\textwidth}{!}{%
\begin{tabular}{l c c c c c c}
\toprule
\hd{Capability} & \hd{Truman} & \hd{Neubrander} & \hd{Slurk/} & \hd{Deliberate} & \hd{OASIS} & \hd{VIRENA} \\
 & & \hd{et al.} & \hd{hivelab} & \hd{Lab} & & \\
\midrule
Feed-based interfaces        & \cmark & \cmark & \xmark & \xmark & \cmark & \cmark \\
Chat-based interfaces        & Limited & \xmark & \cmark & \cmark & Limited & \cmark \\
Replicates specific platforms & \xmark & \xmark & \xmark & \xmark & \xmark & \cmark \\
Live multi-user interaction  & \xmark & \xmark & \cmark & \cmark & \xmark\textsuperscript{a} & \cmark \\
LLM-powered agents           & \xmark & \cmark & \cmark & \cmark & \cmark & \cmark \\
Human participants           & \cmark & \cmark & \cmark & \cmark & \xmark & \cmark \\
Manipulable affordances      & \xmark & \cmark & \xmark & Limited & \xmark & \cmark \\
Configurable moderation      & Limited & \xmark & \xmark & Limited & \xmark & \cmark \\
No-code setup                & \xmark & \xmark & \xmark & \cmark & \xmark & \cmark \\
\bottomrule
\end{tabular}}
\caption{Comparison of platform capabilities. \textsuperscript{a}Agents only.}
\label{tab:comparison}

\end{table}

\FloatBarrier

\section{Platform Overview}\label{sec:overview}

\subsection{Design Principles}

VIRENA is built on five principles:

\tb{Realism}Interfaces follow the design and interaction patterns of the real platforms so that participants engage naturally.

\tb{Control}Researchers manipulate content, timing, participants, agents, and moderation while holding everything else constant.

\tb{Accessibility}Simulations are built in a visual interface without code.

\tb{Transparency}Every setting of an experiment is explicit in the interface, so conditions can be documented and replicated.

\tb{Ethics}All interactions happen in closed environments, and VIRENA collects only the data that the study itself generates. Data are hosted in Switzerland under Swiss data protection standards. 

\subsection{Supported Platforms}

VIRENA offers six platform interfaces in two categories: feed-based platforms (Instagram, Facebook, Reddit, X) and chat-based platforms (WhatsApp, Messenger). Researchers choose the platform when they create an experiment, and it applies to all scenarios of that experiment. Section~\ref{sec:platforms} describes the options of each platform.

\subsection{Key Concepts}

\tb{Organisation}The workspace of a research group. It holds experiments, agents, moderators, and author identities.

\tb{Experiment}A study on one platform. It sets the duration, the waiting room, and the participant limits.

\tb{Scenario}One condition of an experiment. It holds the content (posts or messages with replies), the agents, and the moderators of that condition. The results list scenarios as conditions.

\tb{Instance}One run of the experiment. Participants (one or several) are assigned to an instance and experience one scenario.

\tb{Author identity}The public persona (name and picture) behind a post, reply, agent, or moderator.

\tb{Agent}An LLM-powered participant with a persona.

\tb{Moderator}A rule that detects content and reacts to it.

\section{User Guide}\label{sec:guide}

This section follows the order in which a study is built and run.

\subsection{Navigation}

The left sidebar has four entries: \textit{Experiments}; \textit{Agents \& Identities} with the sub-pages \textit{Agents}, \textit{Moderators}, and \textit{Author identities}; and \textit{Administration}. A switcher at the top selects the organisation.

\textit{Experiments} lists all experiments of the organisation with their status, platform, duration, participant limits, waiting duration, and maximum number of instances. The list can be searched by title, filtered by status (all, active, inactive) and platform, sorted (for example by last modification), and shown as a table or as cards. Experiments can be archived and found again under the \textit{Archived} tab.

\shot{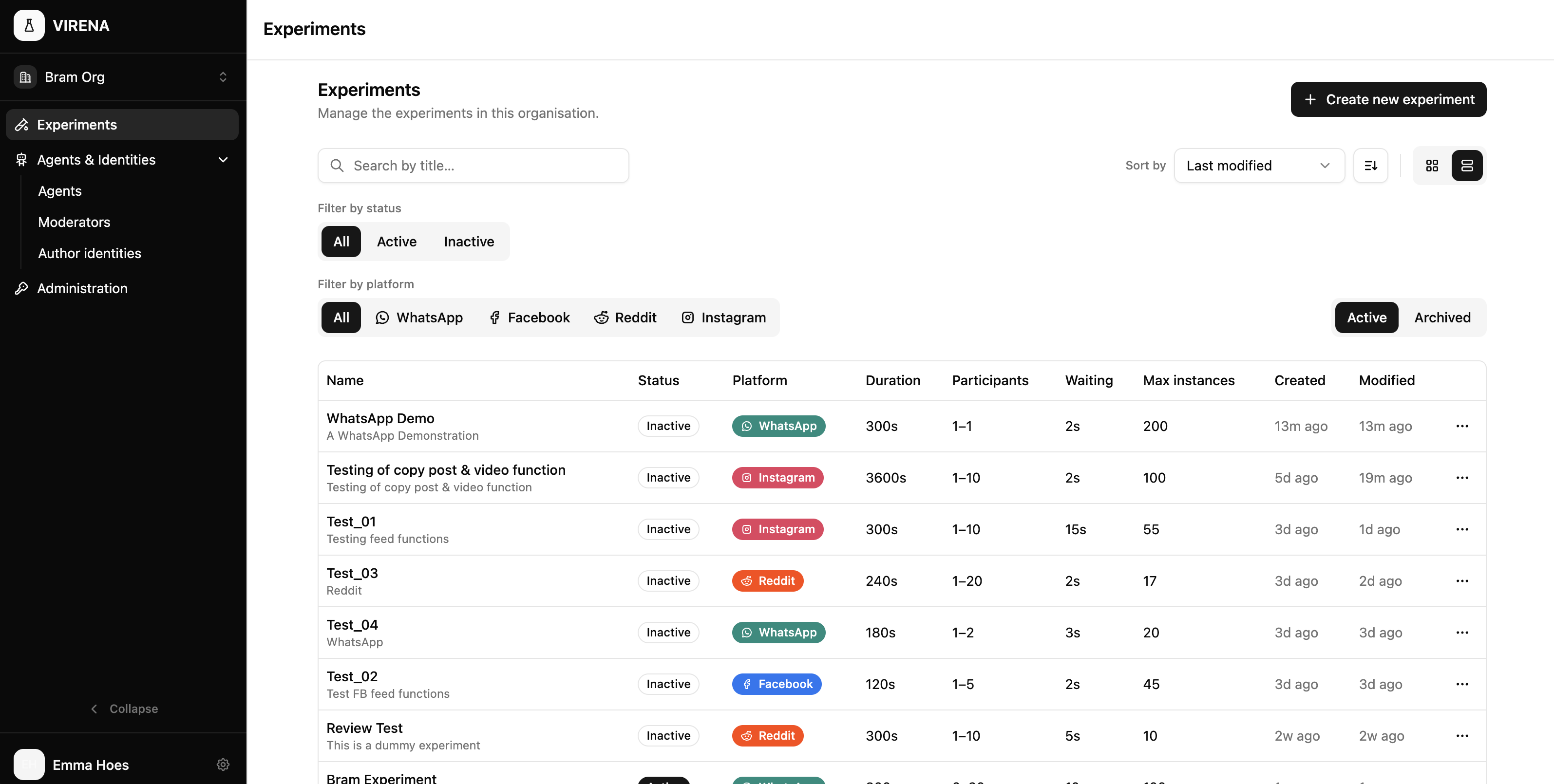}{The experiment list with the navigation sidebar.}{fig:dashboard}{0.94}

\subsection{Step 1: Create an Experiment}

\textit{Create new experiment} opens a form with these fields:

\begin{itemize}
  \item \textbf{Title} and \textbf{Description}. A switch controls whether participants see them.
  \item \textbf{Platform}: Instagram, Facebook, Reddit, X, WhatsApp, or Messenger. The platform is fixed once the experiment is created.
  \item \textbf{Max instances}: the number of runs. If each run holds one participant and 100 participants are needed, set 100. If runs are groups, set the number of groups.
  \item \textbf{Post order}: \textit{Fixed order} (the order set by the researcher) or \textit{Randomized} (a random order for each participant, identical when that participant reconnects). A scenario can override this setting.
  \item \textbf{Duration (seconds)}: the length of a run.
  \item \textbf{Waiting duration (seconds)}: the time the waiting room stays open.
  \item \textbf{Min participants} and \textbf{Max participants}: the limits per instance. Setting both to 1 gives individual sessions.
\end{itemize}

\shot{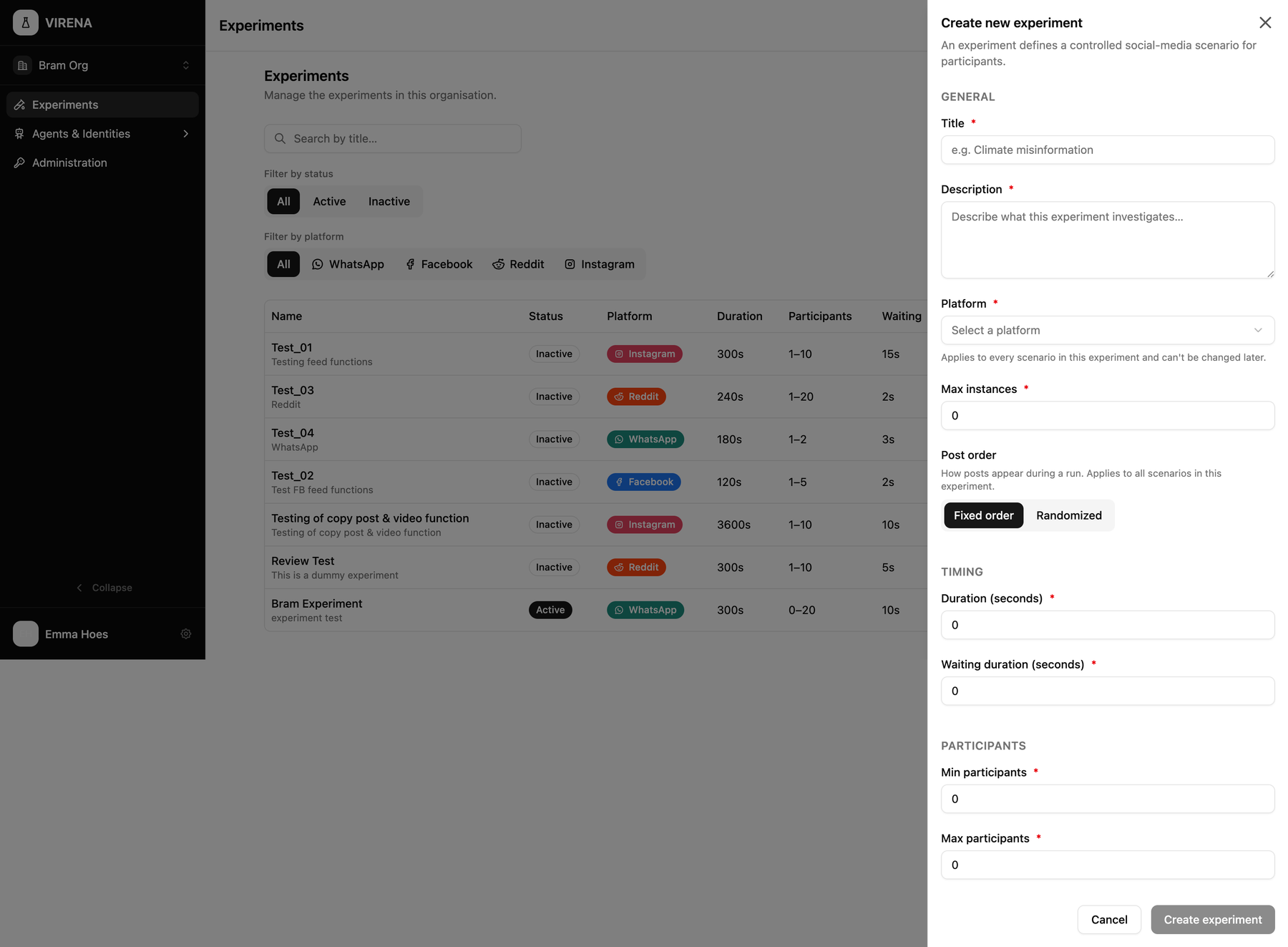}{The Create new experiment form.}{fig:create-exp}{0.62}

\subsection{Step 2: Build Scenarios}

Each scenario has a panel with its content (left) and its settings (right). \textit{Add scenario} creates a new condition.

\shot{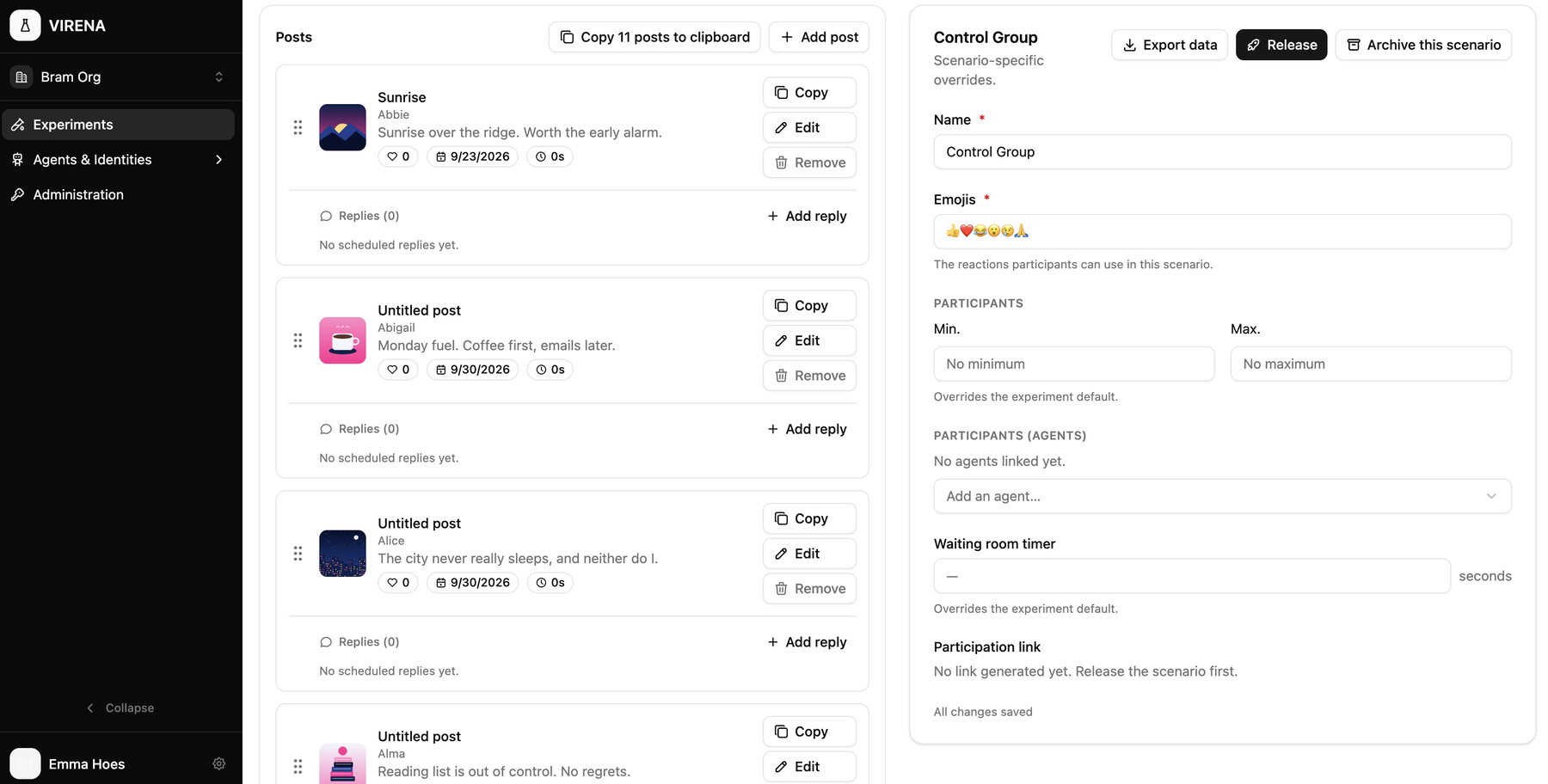}{A scenario with the Posts panel (left) and the scenario settings (right).}{fig:scenario}{0.94}

\tb{Scenario settings}

\begin{itemize}
  \item \textbf{Name}: the label of the condition in the dashboard and in the results.
  \item \textbf{Emojis}: the reactions available to participants in this scenario. Each platform uses the set in its own way (Section~\ref{sec:platforms}).
  \item \textbf{Participants (min, max)}, \textbf{Waiting room timer}, and \textbf{post order}: values for this scenario that override the experiment defaults.
  \item \textbf{Participants (agents)}: the agents that take part in this scenario.
  \item \textbf{Moderators}: the moderation rules that apply to this scenario (Section~\ref{sec:moderation}).
  \item \textbf{Release}: makes the scenario available and generates its \textbf{Participation link}. % TODO: confirm what Release does exactly
  \item \textbf{Participation link}: a link to this scenario alone. It serves designs in which randomization at the individual level is handled by external software such as Qualtrics, which sends each participant to the link of the assigned scenario.
  \item \textbf{Export data}, \textbf{Stop runs}, and \textbf{Archive this scenario}. Archived scenarios can also be deleted.
\end{itemize}

\tb{Posts and messages}\textit{Add post} opens a form. On chat platforms the same list holds messages. The fields are:

\begin{itemize}
  \item \textbf{Title}: the name of the post. It appears in the feed where the platform has titles (Reddit) and names the post in the data export.
  \item \textbf{Author}: an author identity.
  \item \textbf{Caption}: the text of the post or message.
  \item \textbf{Image} (up to 2~MB) or \textbf{Video URL} (YouTube, including Shorts, or Vimeo). Instagram posts accept several images as a carousel.
  \item \textbf{Engagement}: the starting number of likes, votes, reactions, or shares that the platform displays.
  \item \textbf{Date}: the date shown on the post, either fixed or relative to the start of the experiment (for example 48 hours earlier).
  \item \textbf{Delay (seconds after run start)}: when the post appears. A delay of 0 shows the post from the start.
  \item \textbf{Location}: the location tag, for platforms that display one.
\end{itemize}

In fixed order, a drag handle on each post changes its position.

\tb{Replies and reactions}Under each post, \textit{Add reply} opens a form with an author, a type (\textit{message} or \textit{reaction}), the content, the item it answers (the post or another reply, which creates nested threads), a delay, and a date. A reaction uses an emoji from the scenario's emoji set as its content. Nested replies are timed after their parent, and all replies are timed at or after the post. \textit{Generate comments via agents} (in the post form) lets the agents linked to the scenario write a chosen number of comments for a post, which the researcher can then edit. % TODO: confirm the "number of comments" option

\tb{Copying content}\textit{Copy} on a post copies that post, and \textit{Copy posts to clipboard} copies all posts of the scenario together with their replies, reactions, and engagement counts. In an empty scenario, \textit{Paste posts from clipboard} adds the copied posts. This works within an experiment and across experiments, so researchers build a stimulus set once and reuse it.

\shot{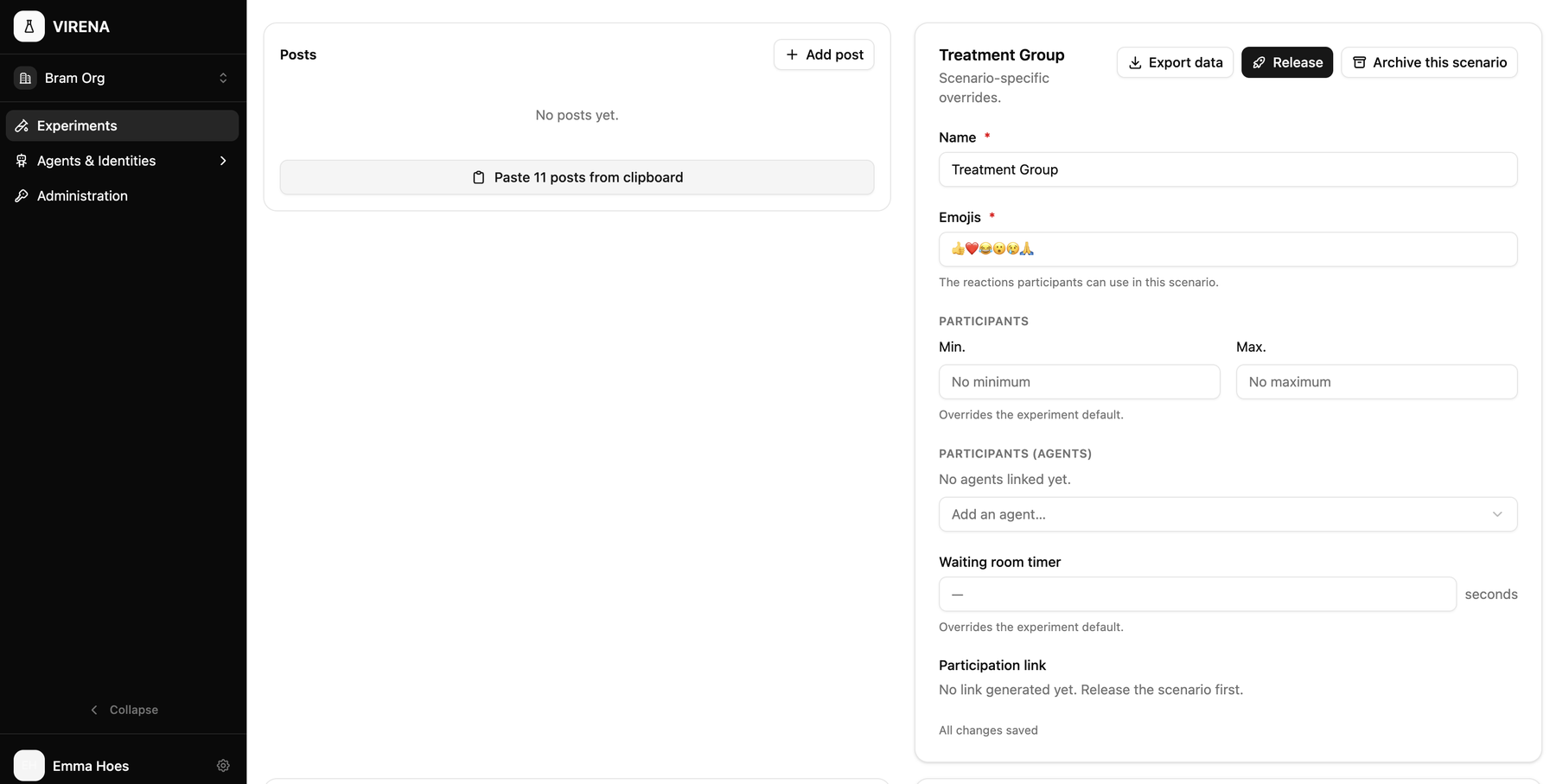}{An empty scenario with the option to paste copied posts.}{fig:copy}{0.94}

\subsection{Step 3: Agents and Identities}

\tb{Author identities}An identity is a public persona (name and picture). The same identity can serve scripted posts, agents, and moderators. Pictures come from a library with gender metadata for balanced or controlled assignment.

\tb{Agents}The \textit{Agents} page holds the agent library, which can be searched and filtered by behaviour (bot or human) and activation. \textit{Create agent} opens a form with these fields:

\begin{itemize}
  \item \textbf{Agent name}: an internal name. Participants see the identity.
  \item \textbf{Identity}: an existing identity or a new one.
  \item \textbf{Persona / system prompt}: the persona, views, and style of the agent, from a single sentence to a detailed backstory.
  \item \textbf{LLM key}: the researcher's own API key for the chosen model provider.
  \item \textbf{Behaviour}: \textit{Human} or \textit{Bot}. This adds a base prompt that shapes the agent's language so that it sounds like a typical human user or like an openly artificial one. The persona prompt is added on top.
  \item \textbf{Activated by}: the events that trigger the agent's replies, such as a participant post.
  \item \textbf{Response delay (seconds)} and \textbf{Randomize delay}: the agent waits the set time, or a random time up to it.
\end{itemize}

\tb{Controlling cost}Agents call a language model for every reply, so costs grow with the number of replies. When agents are set to react to other agents, researchers set a spending or token limit with the model provider, for example by using a dedicated API key for each study. Most providers offer such limits per key or per project. A response delay slows down exchanges between agents.

\shot{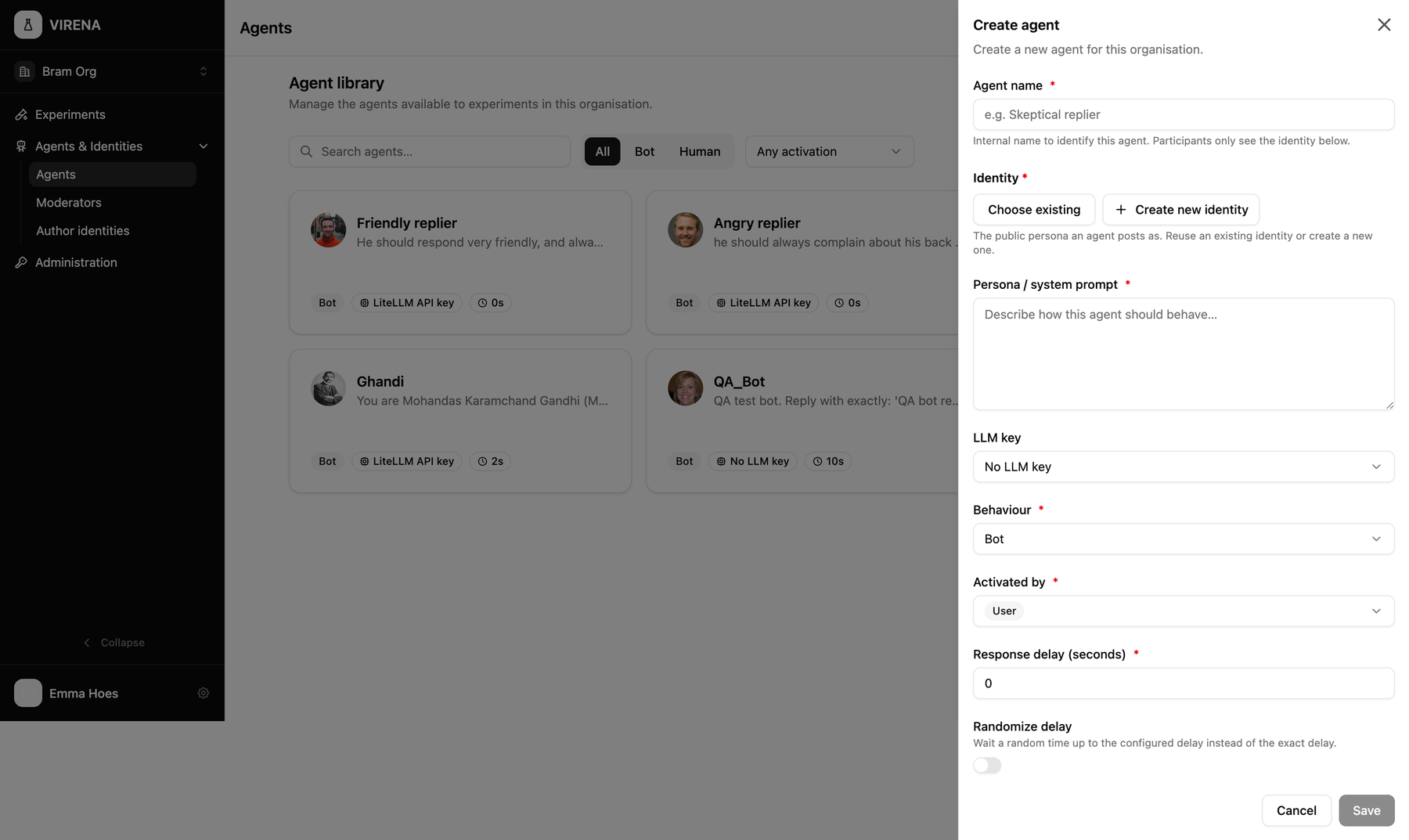}{The Create agent form.}{fig:agent}{0.78}

\subsection{Step 4: Moderation}\label{sec:moderation}

The \textit{Moderators} page holds the moderator library, which can be filtered by rule type. \textit{Create moderator} opens a form with these fields:

\begin{itemize}
  \item \textbf{Identity}: the persona that attributes flags and posts popup warnings.
  \item \textbf{Rule type}: \textit{Keywords}, \textit{Regular expression}, or \textit{AI prompt}. An AI prompt rule lets an LLM classify content according to the researcher's prompt.
  \item \textbf{Keywords}, pattern, or prompt: the content of the rule, depending on the rule type.
  \item \textbf{Reaction}: \textit{Flag}, \textit{Delete}, or \textit{Popup warning}.
  \item \textbf{Warning / reason text}: shown to the participant in a popup warning and stored as the reason of a flag.
  \item \textbf{LLM key}: for AI prompt rules.
  \item \textbf{Activated by}: the message authors the moderator watches (participants, agents, or both).
  \item \textbf{Ban threshold}: the number of violations after which an author is banned. 0 means no ban.
  \item \textbf{Reaction delay (seconds)} and \textbf{Randomize delay}: the time until the moderator reacts, which allows studies of moderation speed.
\end{itemize}

A popup warning appears as an overlay on the screen of the participant who triggered the rule. Researchers add moderators from the library to a scenario, so each scenario can have its own moderation. Banned participants appear in the results.

\shot{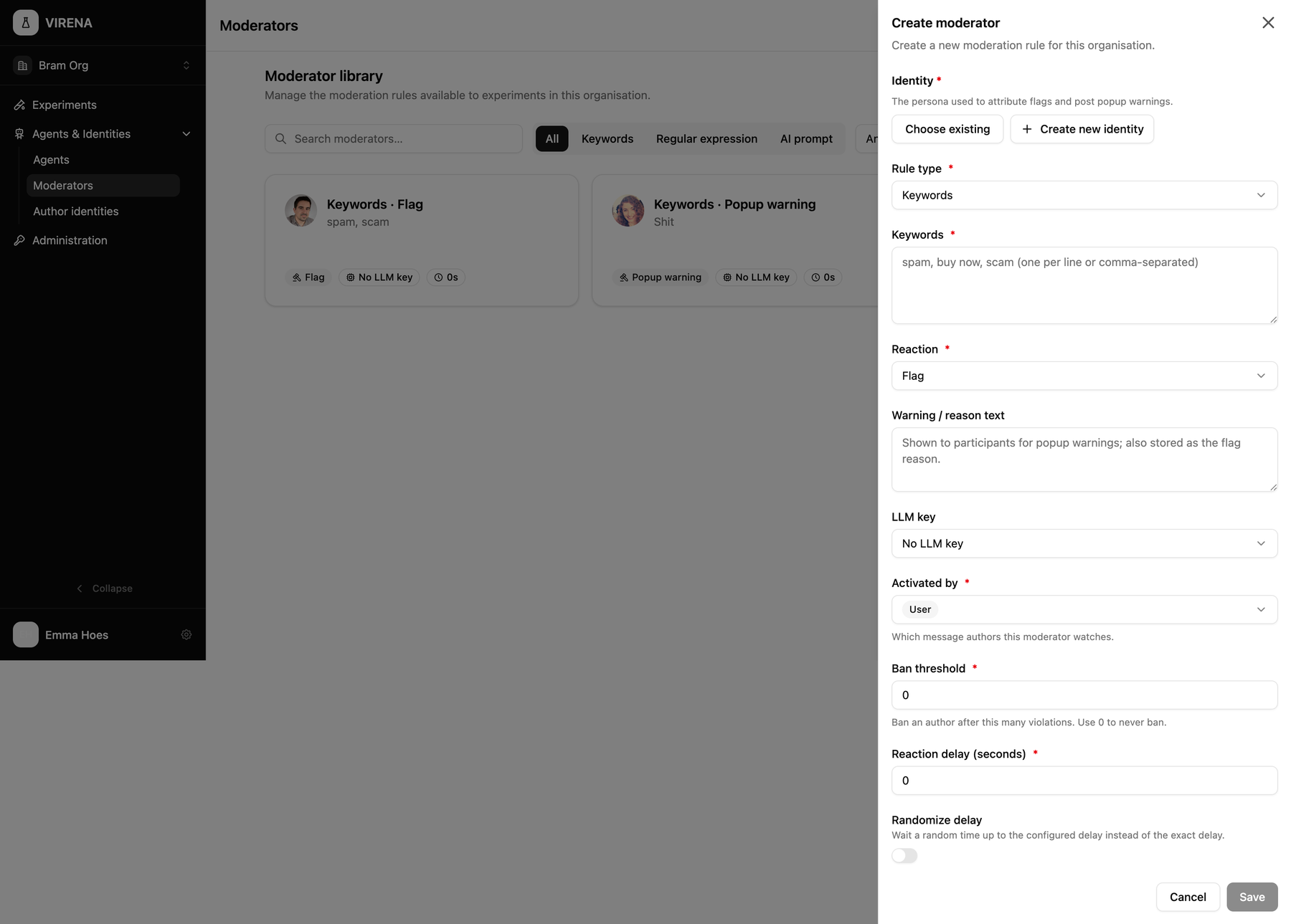}{The Create moderator form.}{fig:moderation}{0.62}

\subsection{Step 5: Recruit and Run}

\tb{Participation links}Each experiment has one participation link, generated when the experiment is released. Participants who open it enter the waiting room and are assigned to a scenario at random. Each scenario also has its own link (Step 2). A link accepts an external participant ID as a URL parameter (for example the Prolific ID), which VIRENA stores and exports. % TODO: confirm where the experiment-level link and its release appear

\tb{Waiting room and assignment}Participants wait in a room that shows how many people have arrived and how much waiting time is left. A run starts as soon as the maximum number of participants is reached. If the maximum is not reached, the run starts when the waiting duration expires, provided that the minimum number of participants has arrived. If the timer expires below the minimum, participants see an end page that states that the requirement was not met.

The group in a room is assigned together to one scenario, chosen at random so that scenarios fill evenly. A new room opens for the next group, and the process repeats until the maximum number of instances is reached. This is randomization at the group level, and it also works for single participants.

\tb{During a run}Participants see the platform interface with a countdown timer and their own name in the header. Scripted content appears at its delay times, agents respond to their triggers, and moderators act on matching content. When the duration ends, participants continue to the redirect URL, for example a follow-up survey, or to an end page.

\tb{Monitoring}\textit{Live results} shows, for each condition, the number of runs (running and done) and of participants assigned, completed, in progress, dropped out, and banned. \textit{Refresh} updates the table. \textit{Stop runs} ends the runs of one scenario, and \textit{Stop all runs} ends all runs. Participants whose run was stopped are recorded as stopped.

\subsection{Step 6: Export Data}

\textit{Export data} produces files for one scenario or for the whole experiment. Files are UTF-8 CSV (and JSON) and include a column for the scenario (condition). % TODO: confirm formats

\begin{itemize}
  \item \textbf{Participants}: external ID, scenario, completion status (completed, dropped out, banned, stopped), and timestamps.
  \item \textbf{Content}: every post, comment, message, and reply with its author, its source role (participant, agent, moderator, script, system), the item it answers, and the time it appeared in the run. Scripted content carries its run time, and an exposure log records when each participant saw each item.
  \item \textbf{Engagement}: all votes, reactions, and shares as events with timestamps, including removals, so the final state and the full sequence can be rebuilt.
  \item \textbf{Behavior}: clicks and attention time per post. % TODO: define precisely how clicks and attention are computed
  \item \textbf{Moderation}: flags, deletions, popups, and bans with the rule that triggered them.
\end{itemize}

Each post carries a stable post ID and the name given in the post form (the post number if no name is given), which keeps the columns of randomized feeds easy to match.

\FloatBarrier

\section{Platform-Specific Interfaces}\label{sec:platforms}

\begin{table}[!htbp]
\centering
\small
\renewcommand{\arraystretch}{1.25}
\begin{tabular}{l l p{8.5cm}}
\toprule
\hd{Platform} & \hd{Type} & \hd{Interaction options} \\
\midrule
Instagram & Feed & Likes, comments, shares; images, carousels, video \\
Facebook  & Feed & Reactions, comments, shares; images, video \\
Reddit    & Feed & Upvotes and downvotes, nested comments, post titles \\
X         & Feed & Likes, reposts, replies \\ % TODO: confirm
WhatsApp  & Chat & Messages, reactions to a message, direct reply to a message \\
Messenger & Chat & Messages, reactions to a message, direct reply to a message \\ % TODO: confirm
\bottomrule
\end{tabular}
\caption{Platform interfaces and their interaction options.}
\label{tab:platforms}
\end{table}

\begin{figure}[H]
\centering
\begin{subfigure}[t]{0.48\textwidth}
  \centering
  \framedimg{0.96\linewidth}{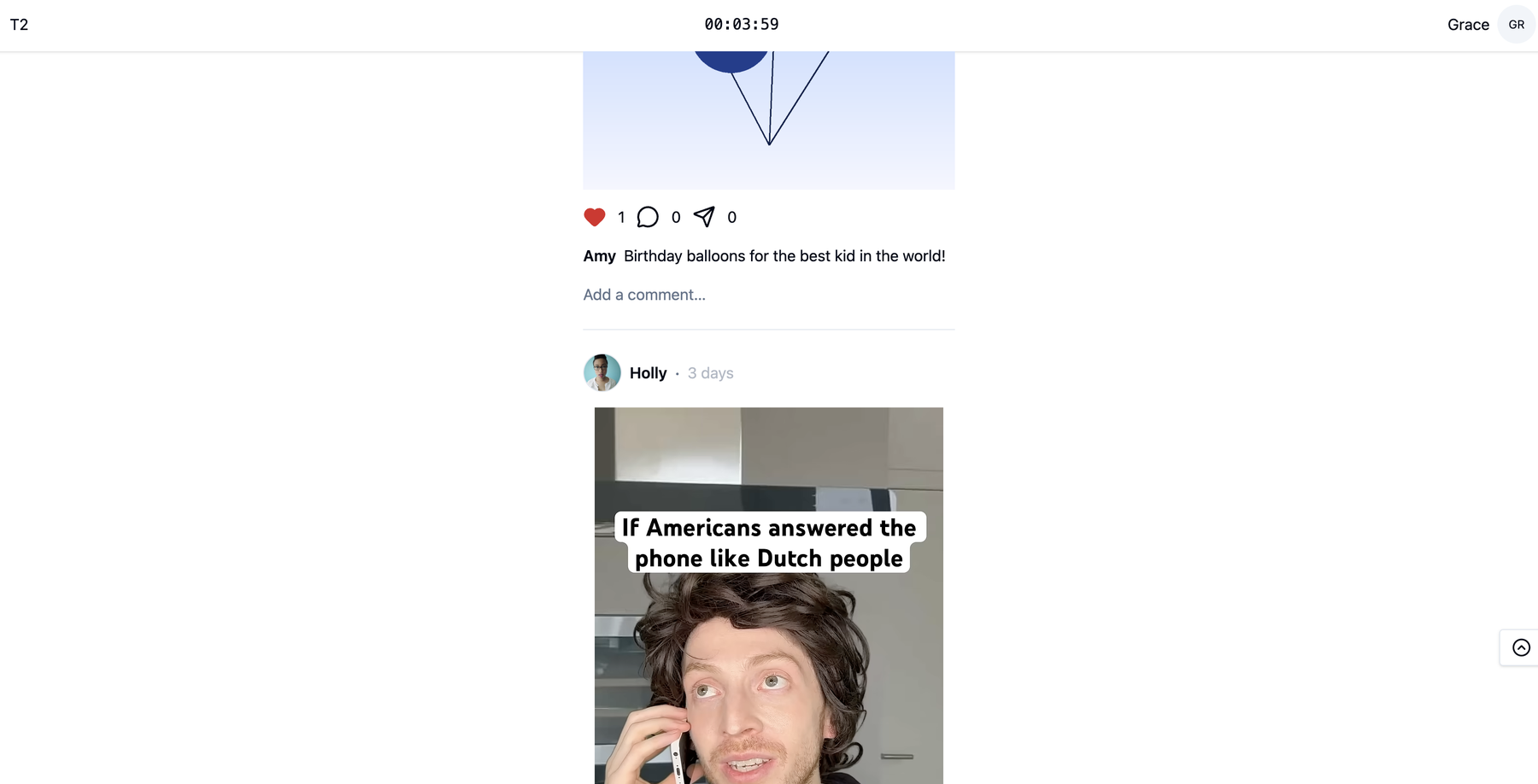}
  \caption{Instagram feed}
\end{subfigure}\hfill
\begin{subfigure}[t]{0.48\textwidth}
  \centering
  \framedimg{0.96\linewidth}{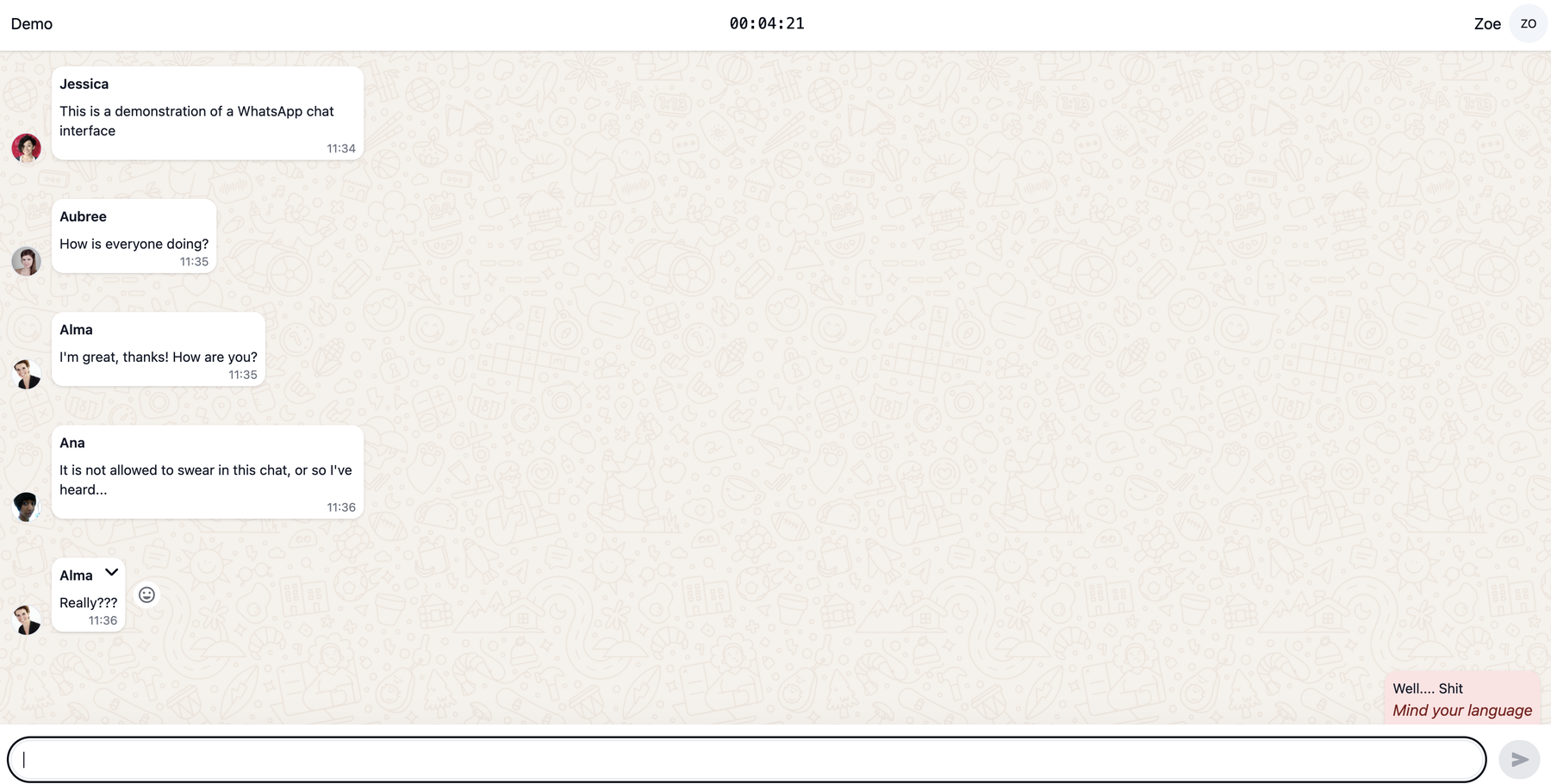}
  \caption{WhatsApp chat}
\end{subfigure}
\caption{The participant view on a feed platform and on a chat platform.}
\label{fig:participant}
\end{figure}

The setup forms follow the platform. A Reddit scenario asks for votes and titles, a Facebook scenario for reactions, and an Instagram scenario for likes, shares, and carousel images. Chat scenarios work with messages that have a sender, a time, and reactions. Participants see the familiar interface of the chosen platform (Figure~\ref{fig:participant}).

\section{Use Cases}\label{sec:usecases}

The following examples illustrate some of the ways VIRENA can be used. They are starting points, not an exhaustive list.

\subsection{Research Applications}

\tb{Misinformation studies}Present controlled false claims in realistic settings, let agents spread or counter them, and test accuracy prompts, source labels, or social correction.

\tb{Moderation research}Compare keyword-based, AI-based, and no moderation, and compare deletion, warnings, and flags, with instant or delayed timing.

\tb{Political communication and deliberation}Test message framing, group composition, and agents with different viewpoints among real participants.

\tb{Group dynamics}Study consensus formation, polarization, and bystander behavior as they unfold.

\tb{AI persuasion}Test how AI-generated content changes attitudes and behavior, with agents set to behave as ``human'' or ``bot''.

\tb{Cross-platform designs}Run a feed scenario and then a chat scenario to study how dynamics carry over between platforms.

% TODO: add pilot studies once they are public

\subsection{Educational Applications}

\tb{Digital literacy}Students experience algorithmic curation, viral spread, content moderation, and AI-generated content in a closed environment.

\tb{Safe practice}Students encounter harassment, misinformation, and polarized debates without real-world consequences.

\subsection{Public and Political Applications}

\tb{Message testing}Test how messages perform with representative audiences before public release.

\tb{Civic engagement}Test messages designed to encourage political participation, volunteering, or civic action.

\tb{Training}Practice responding to hostile online interactions in realistic but consequence-free environments.

\section{Discussion}\label{sec:discussion}

\subsection{Contributions}

VIRENA makes several contributions to the research infrastructure for studying digital communication:

\tb{Integration of capabilities}Platform-specific interfaces, live multi-user interaction, AI agents, and configurable moderation come together in one tool.

\tb{Moderation as a research variable}Researchers can manipulate detection methods, response types, and timing.

\tb{Accessibility}No-code configuration removes technical barriers that have limited who can conduct social media research.

\tb{Reproducibility}A scenario fully specifies a condition, including content, agents, and moderation. Scenarios and their content can be copied and shared, so other researchers can replicate or extend a study.

\tb{Simplified ethical review}Studies run in closed environments with no connection to real accounts, so review boards avoid the complications of real-platform experiments, such as uncontrolled exposure to harmful content or risks to participants' digital footprints.

\subsection{Limitations}

\tb{Ecological validity}Simulated environments cannot fully replicate real social media. Participants may behave differently knowing they are in a study, and effects may not generalize to real platforms.

\tb{AI agent realism}Subtle cues may reveal that an agent is artificial, and agent behavior depends on the prompt and the model.

\tb{Temporal validity}VIRENA replicates platform designs at a specific point in time. As real platforms change, simulations may become less representative.

\tb{Scale}The number of participants per instance has no hard technical ceiling. Performance under very high concurrency has not been formally stress-tested.

\subsection{Next Steps}

Table~\ref{tab:roadmap} lists planned features. Further priorities will follow feedback from pilot collaborations and evolving research needs.

\begin{table}[!htbp]
\centering
\small
\renewcommand{\arraystretch}{1.25}
\begin{tabular}{l p{10cm}}
\toprule
\hd{Feature} & \hd{Description} \\
\midrule
Video platforms          & TikTok, YouTube, Instagram Reels with autoplay \\
Content import           & Pre-populate feeds and chats with real-world content from existing platforms via a URL or reference, including media, captions, and engagement metrics \\
Algorithmic feeds        & Chronological, engagement-based, and personalized ranking \\
Rich content types       & Polls and stickers \\
Multi-language interface & Support for non-English platform interfaces \\
\bottomrule
\end{tabular}
\caption{Planned features.}
\label{tab:roadmap}
% TODO: confirm roadmap
\end{table}

\FloatBarrier

\section{Conclusion}

VIRENA gives researchers, educators, and organizations a realistic but controlled environment for studying digital communication. It combines platform replicas, live human interaction, AI agents, and configurable moderation, and it requires no programming. As access to real platforms continues to shrink, purpose-built research environments become essential.

VIRENA is under active development at the University of Zurich. Researchers, educators, and organizations interested in demos, pilot studies, or collaborative development are invited to contact the authors.

\newpage
\bibliographystyle{apalike}

\end{document}